\documentclass[12pt,cite,epsf,epsfig]{article}
\usepackage{epsfig}

\begin{document}

\author{S. DEV\thanks{dev5703@yahoo.com}
and SANJEEV KUMAR\thanks{sanjeev3kumar@yahoo.co.in}}
\title{Spectral Distortions at Super-Kamiokande}
\date{Department of Physics, Himachal Pradesh University, Shimla 171005, INDIA}
\maketitle

\begin{abstract}
We examine the effect of the rise in the survival probability of the electron
neutrinos with the decrease in the neutrino energy on the recoil electron spectrum
at Super-Kamiokande.
\end{abstract}

The neutral current (NC) measurements at SNO \cite{1} have,
conclusively, established the oscillations of the solar neutrinos
and after the evidence for terrestrial antineutrino disappearance
in a beam of electronic antineutrinos reported by KamLAND
\cite{2}, all other solutions \cite{3,4,5} of SNP can, at best, be
just sub-dominant effects. The solar neutrino experiments have,
already, entered a phase of precision measurements for oscillation
parameters. The completeness of the LMA solution is being questioned
\cite{6} and the scope for some possible sub-dominant transitions is
being explored vigorously \cite{7,8,9,10,11}. The presence of these `new
physics' (NP) effects even at a sub-dominant level will
affect the present determination of the oscillation parameters \cite{11}.
As of now, the exact profile of the survival probability of
electronic neutrinos over the whole energy spectrum remains
unknown as a result of which it is not possible to pin point the
exact mechanism(s) of neutrino flavor conversion or to exclude the
coexistence of the other sub-dominant transitions driven by non-
standard neutrino-matter interactions/ properties. The situation
is complicated by the fact that the transition probability in some
of these non-standard NP scenarios \cite{12} is energy-independent
implying an undistorted solar neutrino spectrum which is
consistent with the Super-Kamiokande and SNO spectral data.

In order to study the effect of the rise in the electron survival
probability on the recoil electron spectrum, we define a quantity $S(T)$ as
\begin{equation}
S(T) =\frac{\left\langle \frac{d\sigma }{dT}\right\rangle _{LMA}}{
\left\langle \frac{d\sigma }{dT}\right\rangle _{SSM}}
=\frac{\int_{E_{\min }}^{E_{\max }}dEf(E)\left[ \frac{d\sigma _{e}}{dT}
P(E)+\frac{d\sigma _{x}}{dT}\left\{ 1-P(E)\right\} \right]
}{\int_{E_{\min }}^{E_{\max }}dEf(E)\frac{d\sigma _{e}}{dT}},
\end{equation}
where $P(E)$ is the LMA survival probability \cite{13} and $f(E)$
is the standard $^8$B neutrino spectrum \cite{14}. The quantity $S(T)$
has the interpretation as the probability of an electron
being scattered with a recoil kinetic energy T and closely
resembles the SK data/SSM ratio with the only difference that the
SK data/SSM ratio is presented for individual energy bins while
$S(T)$ as defined above is a continuous function of recoil
electron energy. The tree level cross-sections for $\nu _{e}e$ and
$\nu _{x}e$ ($x=\mu ,\tau $) scattering have been given in Ref. [15].
In fact, $S(T)$ is the LMA expectation for the
recoil electron spectrum normalized to the standard $^{8}$B
neutrino spectrum and has been plotted as a function of T in Fig.
1 along with the $S(T)$ for energy-independent asymptotic value of
the LMA survival probability for comparison. The two probabilities
differ considerably from each other and this difference could be
as large as 10\% at 5 MeV. This is enough to highlight the role of
spectral distortions in discriminating the LMA suppression
scenario from the energy independent suppression NP scenarios
mentioned above \cite{12}. At the current level of precision, both the LMA
suppression and the energy-independent suppression are consistent
with the SK as well as SNO solar neutrino data. Consequently, one
can  not claim a conclusive confirmation of the LMA solution from
the exclusive study of the high energy region of the solar
neutrino spectrum. It is, therefore, imperative to study
quantitatively the spectral distortions not only to finally
confirm the LMA solution but also to disentangle the possible NP
effects.

The probability that a recoil electron
with true kinetic energy $T_{true}$ will be detected with the observed
kinetic energy $T_{obs}$ is given by
\begin{equation}
r\left( T_{true},T_{obs}\right) =\frac{1}{\sqrt{2\pi }\sigma
_{T}}\exp \left( -\frac{\left( T_{true}-T_{obs}\right)
^{2}}{2\sigma _{T}^{2}}\right)
\end{equation}
and is called the detector response function. The energy-dependent spread, $%
\sigma _{T}$, is of the form
\begin{equation}
\sigma _{T}=\epsilon \sqrt{\frac{T_{true}}{10MeV}}
\end{equation}
so that $\epsilon $ is the energy spread at 10 MeV. For SK,
$\epsilon =1.4$. To account for the finite energy resolution, the
differential cross-section must be folded with the detector
response function, i.e. we must have
\begin{equation}
\left\langle \frac{d\sigma }{dT}\right\rangle \left(
T_{obs}\right) =\int_{0}^{\infty }dT_{true}\left\langle
\frac{d\sigma }{dT}\right\rangle \left( T_{true}\right) r\left(
T_{true},T_{obs}\right)
\end{equation}
in Eq. (1) for $S(T)$ to obtain the observed value of
\begin{equation}
S(T_{obs})=\frac{\int_{0}^{\infty }dT_{true}r\left(
T_{true},T_{obs}\right)
\int_{E_{\min }}^{E_{\max }}dEf(E)\left[ \frac{d\sigma _{e}}{dT_{true}}P(E)+%
\frac{d\sigma _{x}}{dT_{true}}\left\{ 1-P(E)\right\} \right] }{%
\int_{0}^{\infty }dT_{true}r\left( T_{true},T_{obs}\right)
\int_{E_{\min }}^{E_{\max }}dEf(E)\frac{d\sigma _{e}}{dT_{true}}}.
\end{equation}
The quantity $S(T_{obs})$
has been plotted as a function of $T_{obs}$ in Fig. 2 for the LMA
value of survival probability as well as the energy independent
asymptotic value (EIAV) of the survival probability. Also shown
are the same two curves depicted in Fig. 1 assuming perfect energy
resolution. We integrate $S(T_{obs})$ over the total recoil electron
energy $E_{obs}=T_{obs}+m_e$ in the bins of 0.5 MeV. This
integrated normalized spectrum is denoted by $\mathcal{S}$ and has
been plotted in Fig. 3 alongwith the actual 1496 day SK spectrum
normalized to BP04 with statistical errors only \cite{16}.
One can calculate the rise in the LMA value $S_{LMA}(T_{obs})$
relative to the energy independent asymptotic value
$S_{EIAV}(T_{obs})$ at a particular value of $T_{obs}$ by defining
\begin{equation}
R(T_{obs})=\frac{S_{LMA}(T_{obs})-S_{EIAV}(T_{obs})}{S_{EIAV}(T_{obs})}.
\end{equation}
It is instructive to see the variation in $R(T_{obs})$ because of
the energy spread $\epsilon $. Fig. 4 shows $R(T_{obs})$ for
$\epsilon =0$ (perfect energy resolution) and for $\epsilon =1.4$
(finite energy resolution). It is clear that as a result of finite
energy resolution of the detector, $S_{LMA}(T_{obs})$ and,
therefore, $R(T_{obs})$ gets enhanced as compared to the
corresponding values for perfect energy resolution and, again,
this increase is more pronounced at the higher energies. Although,
the curve for $R(T_{obs})$ becomes flatter because of the finite
energy resolution of the detector but the actual value of
$R(T_{obs})$ becomes larger which is an advantage.
One can compare the value of $\mathcal{S}$ in a low
energy bin viz. $\mathcal{S}_L$ with that in a high energy bin
viz. $\mathcal{S}_H$ and find the relative increase in
$\mathcal{S}$ by defining the relative rise-up $\mathcal{R}$ as
\begin{equation}
\mathcal{R}=\frac{\mathcal{S}_L-\mathcal{S}_H}{\mathcal{S}_H}.
\end{equation}
It is important to note that $\mathcal{R}$ defined above is
essentially independent of the flux normalization. Otherwise, it
would have been difficult to directly compare $\mathcal{S}$ with
the experimental data which has to be normalized to the SSM
$^{8}B$ flux which is not known accurately enough. Also, the above
definition is, almost, independent of $\theta _{13}$ since the
factor of $\cos ^{4}\theta _{13}$ cancels in the ratio. A non-zero
$\theta _{13}$ will suppress $S(T)$ by the factor of $\cos
^{4}\theta _{13}$ but the `rise-up' $\mathcal{R}$ will remain
practically unchanged (see Fig. 1). Thus, if the SK spectrum is
found to differ from the LMA spectrum, this conflict cannot be
reconciled even in a 3-flavor framework since the LMA value of
$\mathcal{R}$ is the same in two/three-flavor framework for small
$\theta _{13}$. This is contrary to some assertions made recently
\cite{17} in literature. Moreover, some part of the energy
correlated systematic uncertainties will cancel in $\mathcal{R}$.

The variable $\mathcal{R}$ defined in Eq. (7) can be used as an
observable to quantify the turn-up in the data/SSM ratio at SK at
low energies. It is different from the global observables like the
moments defined in Ref. [18] in the sense that it directly
compares the normalized spectral data at two energy ends of the
spectrum while the moments are global quantities which are not
suitable for this purpose. Moreover, the SK solar neutrino data
is, already, available in small energy bins. For the sake of
illustration, we choose the bins $E=5.5-6.0 MeV$ and $E=13.0-13.5
MeV$ and evaluate $\mathcal{R}$ for these bins to obtain
\begin{equation}
\mathcal{R}=-0.054\pm0.126
\end{equation}
where both the systematic and statistical errors have been
incorporated in the analysis. The rise-up is not significantly
different from zero for these energy bins. The upper bound on
$\mathcal{R}$ is approximately 0.07 (0.20) at $1 \sigma$ ($2
\sigma$) C.L.. This is to be compared with the corresponding LMA
value of rise-up
\begin{equation}
\mathcal{R}_{LMA}^{day}=0.087^{+0.026}_{-0.016}
\end{equation}
for $\Delta m^2=7.9\pm 0.3\times 10^{-5} eV^2$ and $\sin^2
\theta_{12}=0.3^{+0.02}_{-0.03}$ where the earth regeneration
effects have been neglected. However, the earth regeneration
effects can be incorporated in a straight forward manner which
yields
\begin{equation}
\mathcal{R}_{LMA}^{night}=0.070^{+0.024}_{-0.015}.
\end{equation}
The rise-up in the total (day+night) SSM normalized rate is
\begin{equation}
\mathcal{R}_{LMA}=0.078^{+0.025}_{-0.016}.
\end{equation}
This LMA value is compatible with the present experimental value
[Eq. (8)] within 1 $\sigma$ C.L..
The `rise-up' $\mathcal{R}$ in the total (day+night) SSM
normalized rate has been plotted in Fig. 5 as a function of
$\theta _{12}$ for the central and 1 $\sigma$ upper/lower values
of $\Delta m_{12}^{2}$. Since, the rise-up $\mathcal{R}$ becomes
larger for smaller values of $\theta _{12}$, an upper bound on
$\mathcal{R}$ can be used to obtain a lower bound on $\theta
_{12}$ \cite{9}. Fig. 6 depicts the improvement in the
upper bound on $\mathcal{R}$ with increase in statistics and
reduction in systematic errors at Super-Kamiokande. The
confidence levels for the measurement of $\mathcal{R}$ with the
increase in statistics and a projected reduction in systematic errors
at SK have been given in Fig. 7 from which it is clear that with the present
level of accuracy at Super-Kamiokande, a rise-up of $20\%$ can be
measured at $1.1\sigma$ C.L. Not much improvement results even
with the increased statistics with the present level of systematic
errors. The contribution of \textit{hep} neutrinos for the bins we have
examined is very small (about $1.2\%$) for the SSM value of this
flux. However, arbitrarily large values of \textit{hep} flux can
substantially affect the higher energy bins with energy greater
than 14 MeV.

The enhancement in $S(T)$ for smaller values of $T$ can be used to
further constrain the currently allowed neutrino parameter space.
We illustrate this point by plotting the constant $\mathcal{S}$
and $\mathcal{R}$ curves on the ($\Delta m_{12}^{2}$,
$\theta_{12}$) plane.  The quantity $\mathcal{S}$ is obtained by
integrating $S(T)$ over the energy bin of 0.5 MeV centered around
$T=5 MeV$ and the quantity $\mathcal{R}$ is calculated from Eq.
(7) for the energy bins of 0.5 MeV centered around $T=5 MeV$, $15
MeV$. We choose these values for illustrative purposes only. The
constant $\mathcal{S}$ and $\mathcal{R}$ curves on the ($\Delta
m_{12}^{2}$, $\theta _{12}$) plane are shown in Fig. 8 for the
($\Delta m_{12}^{2}$, $\theta_{12}$) parameter space within the
currently allowed LMA region. For $\Delta m_{12}^{2}$ and
$\theta_{12}$, we select the $2 \sigma$ ranges of these quantities
and plot the constant $\mathcal{S}$ curves for
$\mathcal{S}=0.44,0.46,0.48$ (curves with negative slope from left
to right, respectively) and, also, plot the constant $\mathcal{R}$
curves for $\mathcal{R}=0.12, 0.10, 0.08$ (curves with positive
slope from left to right, respectively). It is evident that there
is an increase in $\mathcal{S}$ and decrease in $\mathcal{R}$ with
increasing $\theta_{12}$. Therefore, an accurate measurement of
$\mathcal{S}$ and $\mathcal{R}$ at 5MeV will further constrain the
LMA allowed $\theta_{12}$.  For instance, if $\mathcal{S}$ is
found to be smaller than 0.46 (i.e. $\mathcal{S}\leq 0.46$) and
$\mathcal{R}$ is found to be less than 12\% (i.e. $\mathcal{R}\leq
12\%$), $\theta_{12}$ will, approximately, be within the range
$31^o\le \theta_{12} \le34^o$. In conclusion, the prospects for the
observation of spectral distortions at SK-III, in case it comes up,
appear to be bright \cite{19}. The failure to observe spectral
distortions at SK-III would signal new physics beyond LMA.

The research work of S. D. is supported by Department of Atomic
Energy, Government of India \textit{vide} Grant No. 2004/ 37/ 23/
BRNS/ 399. S. K. acknowledges the financial support provided by
Council for Scientific and Industrial Research (CSIR), India.

\pagebreak

%%%%%%%%%%%%%%%%%%%%%%%%%%%%%%%%%%
\begin{figure}[tb]
\begin{center}
%\vskip 1cm
{\epsfig{file=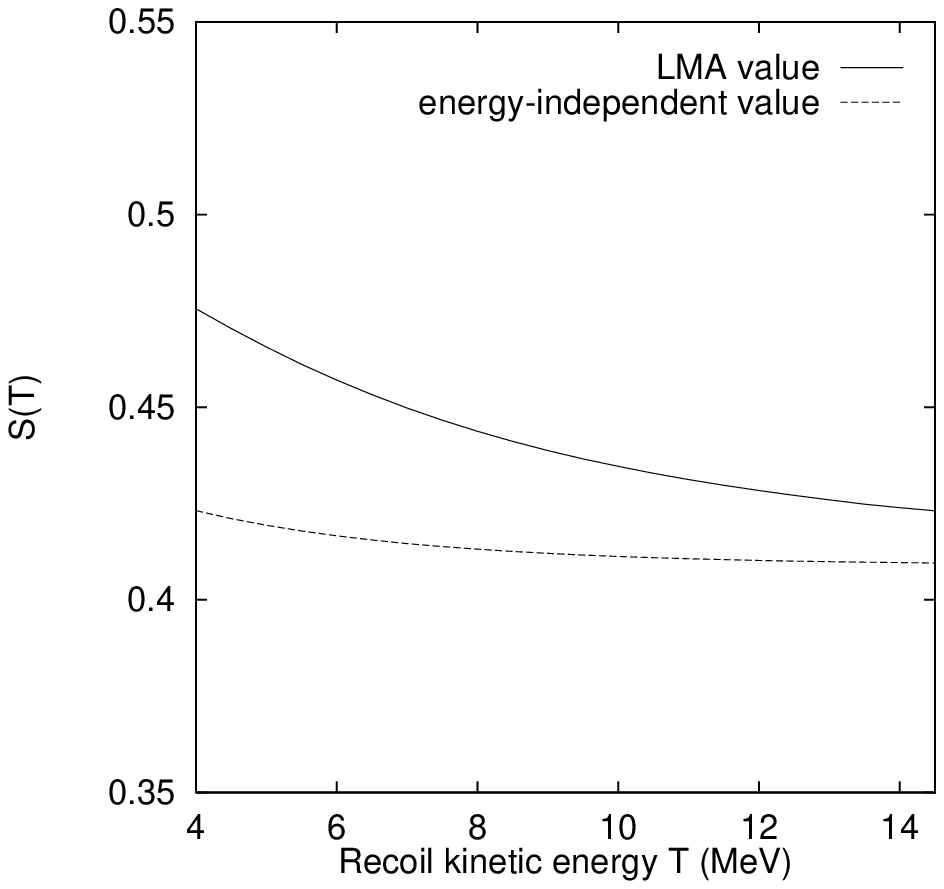, width=8.0cm,
height=7.5cm}\epsfig{file=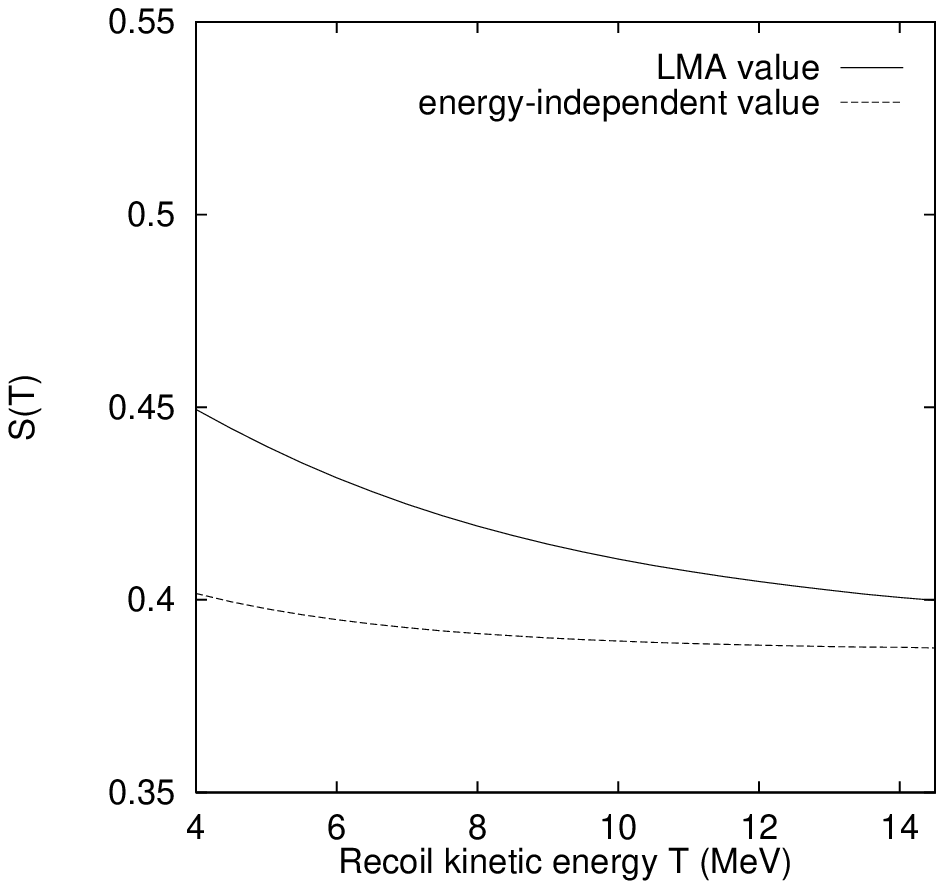, width=8.0cm,
height=7.5cm}}

\end{center}
\caption{S(T) versus T for $\theta_{13}=0^o$ (left panel) and
$\theta_{13}=12.3^o$ (right panel).}
\end{figure}
%%%%%%%%%%%%%%%%%%%%%%%%%%%%%%%%%%

%%%%%%%%%%%%%%%%%%%%%%%%%%%%%%%%%%
\begin{figure}[tb]
\begin{center}
%\vskip 1cm
\rotatebox{0}{\epsfig{file=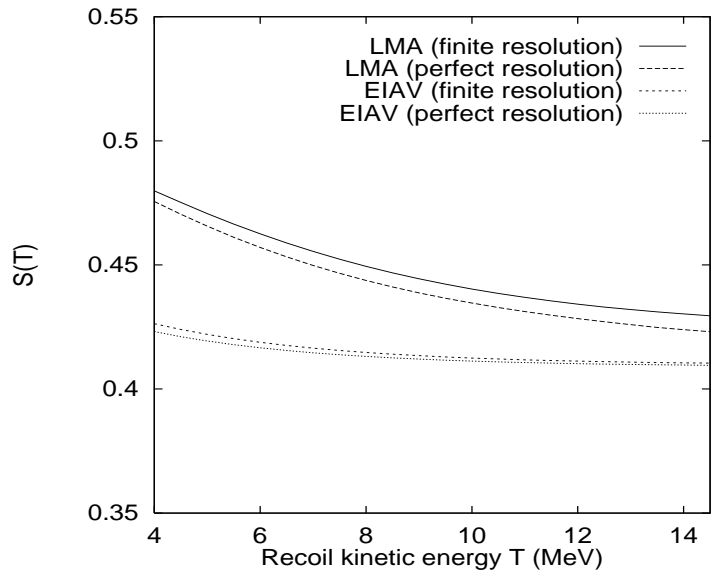, width=10.0cm,
height=7.5cm}}
\end{center}
\caption{Probability folded with detector response function and
neutrino cross-sections as the function of recoil electron kinetic
energy.}
\end{figure}
%%%%%%%%%%%%%%%%%%%%%%%%%%%%%%%%%%

%%%%%%%%%%%%%%%%%%%%%%%%%%%%%%%%%%
\begin{figure}[tb]
\begin{center}
%\vskip 1cm
\rotatebox{0}{\epsfig{file=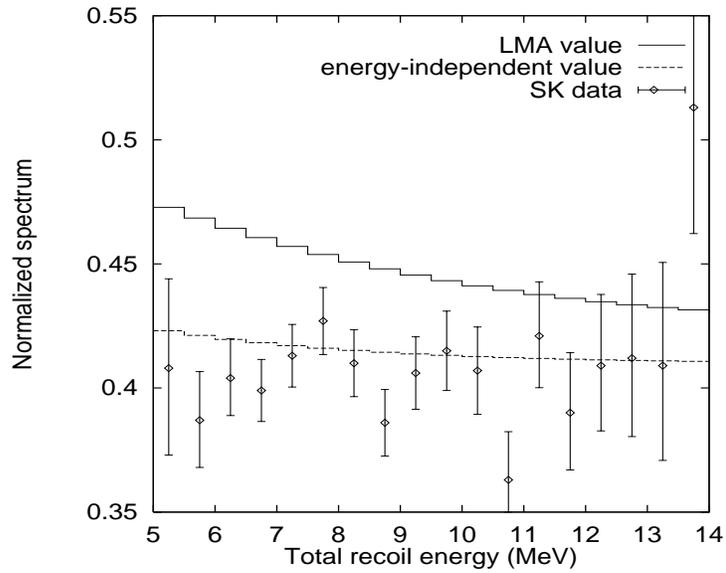, width=10.0cm,
height=7.5cm}}
\end{center}
\caption{The recoil electron spectrum normalized to SSM in the
bins of 0.5 MeV ($\mathcal{S}$). The SK data with statistical
errors is also shown for comparison.}
\end{figure}
%%%%%%%%%%%%%%%%%%%%%%%%%%%%%%%%%%

%%%%%%%%%%%%%%%%%%%%%%%%%%%%%%%%%%%%%%%%%%%%%%%%%%%%%%5
\begin{figure}[tb]
\begin{center}
%\vskip 1cm
\rotatebox{0}{\epsfig{file=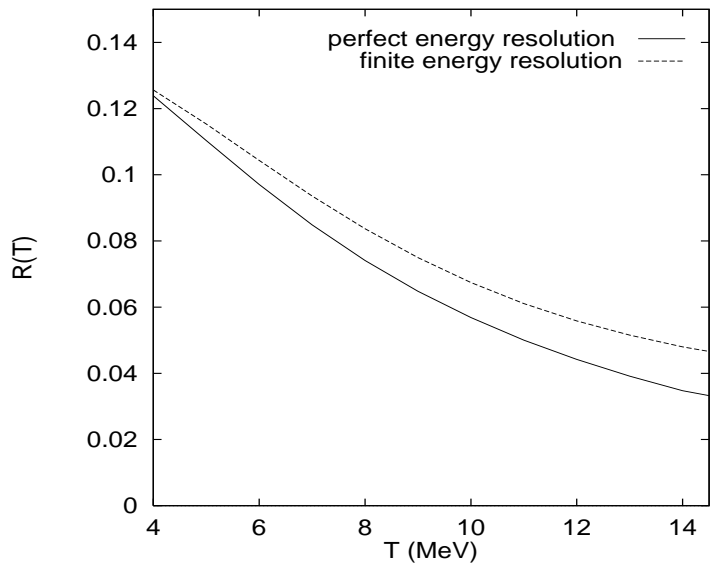, width=10.0cm,
height=7.5cm}}
\end{center}
\caption{$R(E_{obs})$ versus $E_{obs}$.}
\end{figure}
%%%%%%%%%%%%%%%%%%%%%%%%%%%%%%%%%%

%%%%%%%%%%%%%%%%%%%%%%%%%%%%%%%%%%%%%%%%%%%%%%%%%%%%%
\begin{figure}[tb]
\begin{center}
%\vskip 1cm
\rotatebox{0}{\epsfig{file=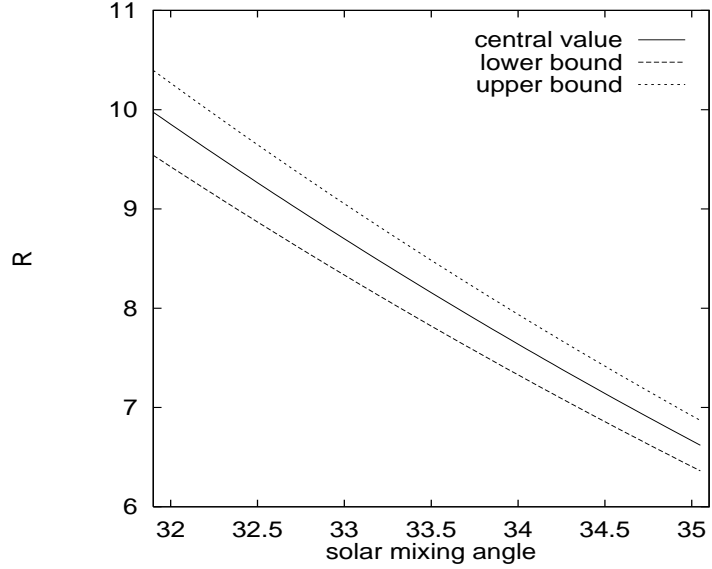, width=10cm, height=7.5cm}}
\end{center} \caption{The rise-up $\mathcal{R}$ in the total (day+night)
SSM normalized rate versus $\theta_{12}$ for the central and 1
$\sigma$ upper/lower values of $\Delta m^2$. The solar mixing
angle $\theta_{12}$ has been varied over its 1 $\sigma$ range.}
\end{figure}
%%%%%%%%%%%%%%%%%%%%%%%%%%%%%%%%%%%%%%%%%%%%%%%%%%%%%

%%%%%%%%%%%%%%%%%%%%%%%%%%%%%%%%%%%%%%%%%%%%%%%%%%%%%%%%%%%%%%%%%%%

\begin{figure}[tb]
\begin{center}
%\vskip 1cm
\rotatebox{0}{\epsfig{file=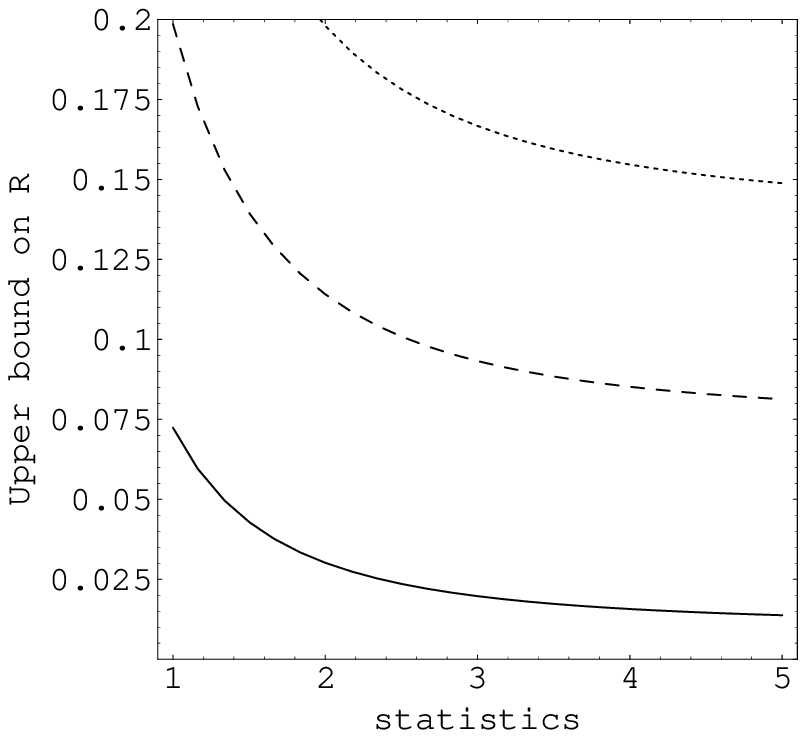, width=7.5cm,
height=7.5cm}}{\epsfig{file=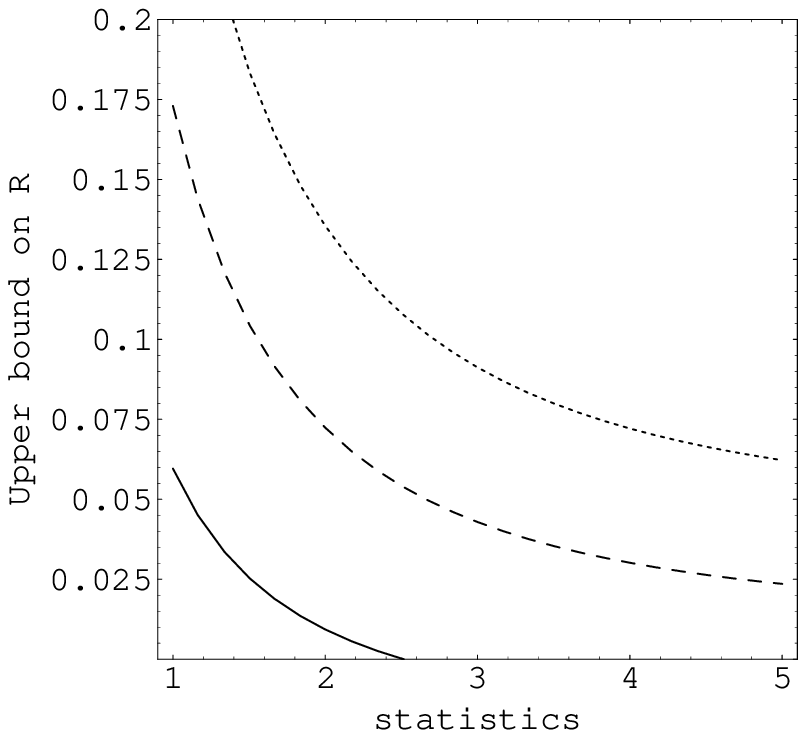, width=7.5cm,
height=7.5cm}}
\end{center}
\caption{Improvements in the upper bound on $\mathcal{R}$ with
increased statistics with present (left panel) and half of the
present (right panel) systematic errors. The solid, dashed and
dotted lines are the upper bounds at 1, 2 and 3 $\sigma$ C.L.}
\end{figure}
%%%%%%%%%%%%%%%%%%%%%%%%%%%%%%%%%%%%%%%%%%%%%%%%%%%%%%%%%%%%%%%%%%

%%%%%%%%%%%%%%%%%%%%%%%%%%%%%%%%%%%%%%%%%%%%%%%%%%%%%%%%%%%%%%%%%%

\begin{figure}[tb]
\begin{center}
%\vskip 1cm
\rotatebox{0}{\epsfig{file=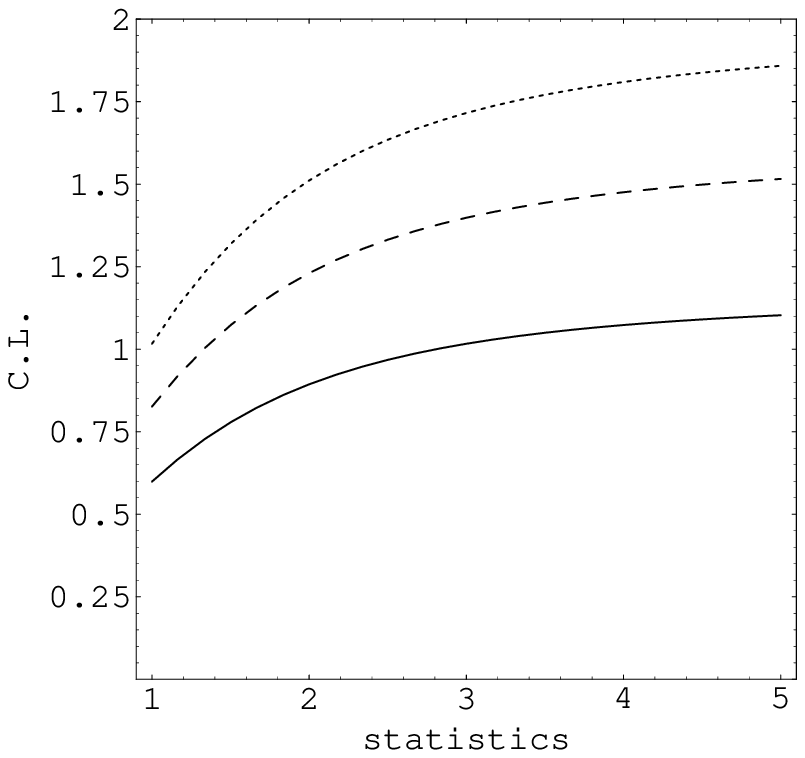, width=7.5cm,
height=7.5cm}} {\epsfig{file=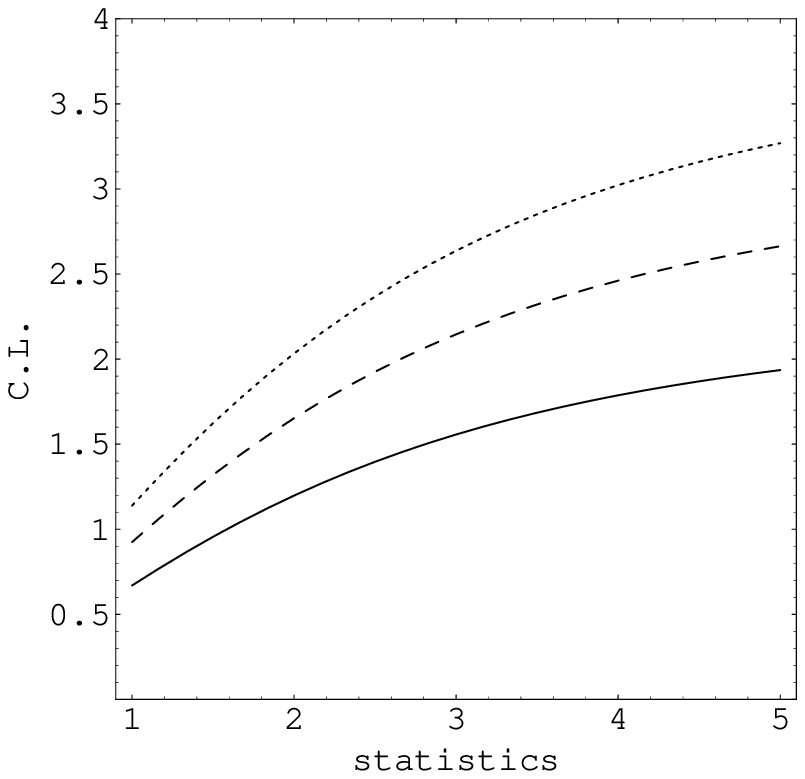, width=7.5cm,
height=7.4cm}}
\end{center}
\caption{(a) Confidence levels for the measurement of
$\mathcal{R}$ with increase in statistics for present systematic
errors (left panel). (b) Confidence levels for the measurement of
$\mathcal{R}$ with increase in statistics for half of the present
systematic errors (right panel). The solid, dashed and  dotted
lines are for $\mathcal{R}=0.10,0.15$ and $0.20$.}
\end{figure}
%%%%%%%%%%%%%%%%%%%%%%%%%%%%%%%%%%%%%%%%%%%%%%%%%%%%%%%%%%%%%%%%%%%

%%%%%%%%%%%%%%%%%%%%%%%%%%%%%%%%%%%%%%%%%%%%%%%%%%%%%%%%%%%%%%%%%%%

\begin{figure}[tb]
\begin{center}
%\vskip 1cm
\rotatebox{0}{\epsfig{file=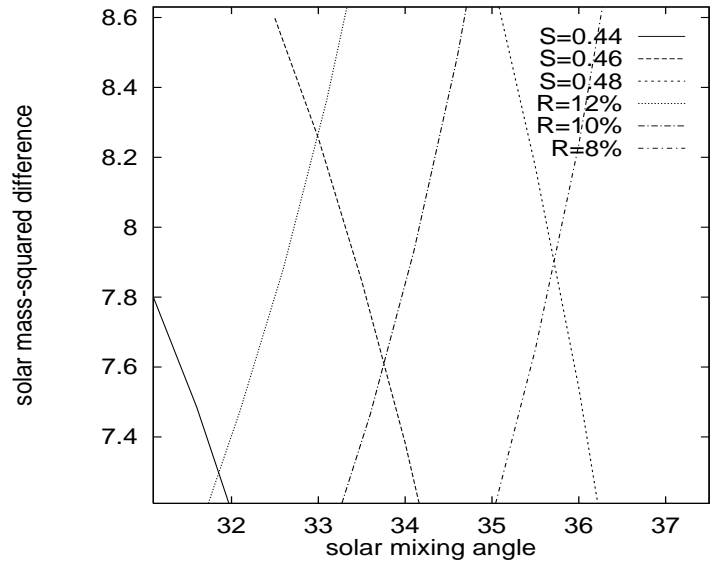, width=10.0cm,
height=7.5cm}}
\end{center}
\caption{Constant $\mathcal{S}$ and $\mathcal{R}$ plots.}
\end{figure}
%%%%%%%%%%%%%%%%%%%%%%%%%%%%%%%%%%%%%%%%%%%%%%%%%%%%%%%%%%%%%%%%%%%%

\end{document}